\documentclass[letter]{aa} 
\usepackage{graphicx}
\usepackage{txfonts}
\newcommand{\degree}{\ensuremath{\mathrm{^\circ}}}
\newcommand{\arcm}{\ensuremath{\mathrm{^\prime}\;}}
\newcommand{\arcs}{\ensuremath{\arcmm\hskip -0.1em\arcmm \;}}
\newcommand{\arcmm}{\ensuremath{\mathrm{^\prime}}}

\newcommand{\dotsec}{\,\rlap{\hbox{$\mathrm{^s}$}}{\hbox{$.$}}\,}

\begin{document}
   \title{A giant radio halo in the low luminosity X-ray cluster Abell 523}

   \author{G. Giovannini \inst{1,2},
           L. Feretti \inst{2}, M. Girardi \inst{3},
           F. Govoni \inst{4}, M. Murgia \inst{4}, V. Vacca\inst{4,5},
           J. Bagchi \inst{6}}

 \institute{(1) Dipartimento di Astronomia, via Ranzani 1, 40127 Bologna, I \\
            (2) Istituto di Radioastronomia-INAF, via P.Gobetti 101,
                40129 Bologna, Italy \\
            (3) Dipartimento di Fisica-Sezione di Astronomia, via Tiepolo 11, 
                34143 Trieste, Italy \\
            (4) Osservatorio Astronomico di Cagliari-INAF,
                Strada 54, Loc. Poggio dei Pini, 09012 Capoterra (Ca), Italy \\
            (5) Dipartimento di Fisica, Universit\`a degli studi di Cagliari, 
                Cittadella Universitaria, 09042 Monserrato (CA), Italy \\
            (6) Inter University Center for Astronomy and Astrophysics (IUCAA),
                 Post Bag 4, Ganeshkhind, Pune 411 007, India
             }

\authorrunning{Giovannini et al.}

   \date{ }

\abstract{Radio halos are extended and diffuse non-thermal 
radio sources present at the cluster center, 
not obviously associated with any individual galaxy. A strong correlation
has been found between the cluster X-ray luminosity and the halo radio power.
}
{We observe and analyze the diffuse radio emission
present in the complex merging structure Abell 523, classified as
a low luminosity X-ray cluster, to discuss its properties in the context of
the halo total radio power  versus X-ray luminosity correlation.
}
{We reduced VLA archive observations at 1.4 GHz to derive a deep radio image
of the diffuse emission, and compared radio, optical, and X-ray data.
}
{Low-resolution VLA images detect a giant radio halo associated with a
complex merging region. The properties of this new halo agree with 
those of radio halos in general
discussed in the literature, but its radio power is about
a factor of ten higher than expected on the basis of the cluster
X-ray luminosity.
}
{Our study of this giant radio source demonstrates that radio halos can 
also be present in clusters with a low X-ray luminosity. Only a few similar 
cases have so far been found . This result suggests that this source represent
a new class 
of objects, that cannot be explained by classical radio halo models.
We suggest that the particle
reacceleration related to  merging processes is  very efficient 
and/or the
X-ray luminosity is not a good indicator of the past merging activity
of a cluster.
} 


   \maketitle
%

\section{Introduction}

Clusters of galaxies are characterized by X-ray emission from a hot 
intra-cluster medium (ICM, $T \sim 2{-}10$ keV). Thermal emission is a common 
property of all clusters of galaxies and has been detected even in poor 
groups as well as in optical filaments connecting rich clusters.
The ICM is believed to be shock heated by merging 
during the structure formation.

Diffuse non-thermal radio sources with steep spectra have been detected
in the central regions of a large number of clusters of galaxies 
(see e.g. \cite{gio09},
\cite{ven08} (and refs therein) and review papers such as 
\cite{fer05}, and \cite{ferra08}).
They are giant radio sources with a spatial extent similar to that of the hot 
ICM, which are called radio halos.
In addition, diffuse radio sources classified as radio relics are
detected in cluster peripheral regions (see e.g. \cite{gio04} and 
\cite{vwe11}).
These sources are not directly associated
with the activity of individual galaxies and are related to physical 
properties of the whole cluster. 
There is a substantial evidence that radio halos are found in 
clusters with significant substructures in X-ray images and
complex gas temperature distributions, both signatures of cluster mergers 
(e.g. \cite{fer99}, \cite{gov04}, \cite{gio09}, \cite{cas10}). 
Classical radio halos show a centrally located and regular radio morphology. 
However, in several cases the radio halos are irregular and elongated as 
in A209, A401, and A2034. 
A few radio halos such as A851 and A2218 are highly asymmetric, 
being located mostly to one side with respect to the cluster center. 
\cite{gio09} interpreted these structures as 
possibly reflecting the geometry of the merger.

\cite{fer00} discussed that the radio power of halos at 1.4 GHz increases 
with the X-ray luminosity of the parent cluster, implying a direct connection 
between
the radio and X-ray plasmas. Since the cluster X-ray luminosity and mass
are correlated, it follows that radio halo power correlates with cluster 
mass (\cite{fer00}; \cite{gov01}).
The strong correlation between the radio halo total power at 1.4 GHz and the 
X-ray cluster luminosity was recently discussed using higher quality 
statistics by
\cite{bru07} and \cite{gio09}, who show that the correlation
is also present for low power radio halos (P$_{1.4}$ $<$ 10$^{24}$ W/Hz and 
L$_x$ $\sim$ 10$^{44}$ erg/s).

\cite{gio09} were the first to discuss whether a radio halo was present
in A1213, an under-luminous X-ray cluster i.e. well outside the above correlation.
A similar case was found in the cluster 0217+70 by \cite{bro11}.
We do not consider radio relics because present models connect
radio halos to turbulence in the ICM, while relic sources
are correlated to the presence of shock waves.
\cite{bro11} discussed whether the low X-ray luminosity in 0217+70
could be due to an underestimate of the X-ray luminosity because of 
the hydrogen column density, which might be patchy at these 
low Galactic latitudes. Therefore, further X-ray observations are required to 
measure the absorption directly. We note that \cite{gio09} 
discussed the peculiarity of the diffuse emission in A1213.

We present new VLA images from archive data of diffuse radio emission, 
found by chance by inspecting NVSS images, in A523, one more 
radio halo in a low X-ray luminosity cluster. 
We present in Sect. 2 the radio data and the 
X-ray and optical information. Discussion and conclusions
are given in Sect. 3.

The intrinsic parameters quoted in this paper are computed for
a $\Lambda$CDM cosmology with $H_0$ = 71 km s$^{-1}$Mpc$^{-1}$,
$\Omega_m$ = 0.27, and $\Omega_{\Lambda}$ = 0.73.
At z =0.10, the luminosity distance is 455 Mpc, and the angular conversion 
factor is 1.82 kpc/arcsec.

\section{Radio images}

A523 was observed  at 1.4 GHz in line
mode in 2005, with the VLA in the C 
(5 hrs) and D (2 hrs) configuration (project AB1180).
Calibration and imaging were performed with the NRAO Astronomical
Image Processing System (AIPS). 
After the editing of bad points, self-calibration 
was applied to both datasets,
to remove residual phase and gain variations. 
  \begin{figure}
   \centering
   \includegraphics[width=7.5cm]{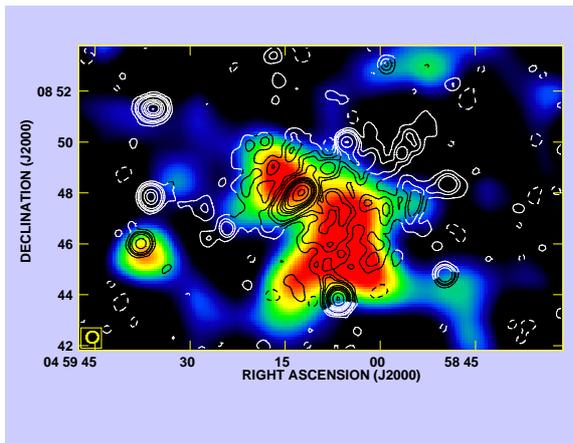}
         \caption{High resolution radio image of A523 at 1.4 GHz
(contours). 
The HPBW is 25.5''$\times$27.1'' in PA -55$^\circ$. The noise level is
0.034 mJy/beam; contours are: -0.1, 0.1, 0.3, 0.5, 1, 3, 5, 7, 10, 15, 30, 
and 50 
mJy/beam.
Colors refer to the X-ray emission detected from the Rosat 
satellite.}
\label{fig:1}
\end{figure}

\begin{figure}
   \centering
   \includegraphics[width=7.5cm]{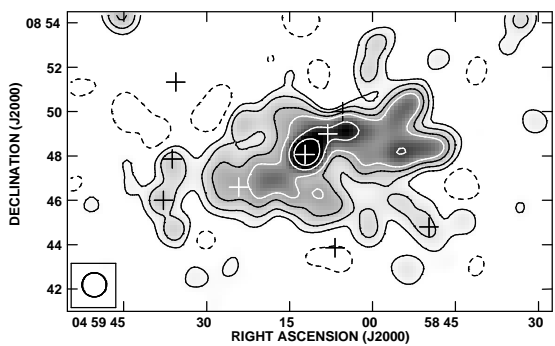}
      \caption{Radio image of the halo in A523 (contours and gray scale) 
after subtraction of 
discrete sources. The HPBW is 65''. The noise level is 0.04 mJy/beam.
Contours are: -0.2, 0.2, 0.5, 1, 1.5, 3, 5 mJy/beam. Crosses indicate the
peak position of subtracted sources.}
         \label{fig:2}
\end{figure}

In Fig. \ref{fig:1}, we show the final image obtained with C configuration 
data and natural weights (contours), superimposed on the X-ray image (colors)
obtained with the Rosat satellite.
The
diffuse emission is also visible at this relatively high angular resolution 
and discrete sources are well separated. 
To obtain a deeper image of the diffuse source, we performed a full resolution
(using uniform weighting) image with C array data, cutting out in addition
the shortest baselines where some flux density from the extended halo was 
present. In this image, only discrete sources are present, and we used it to
select the clean components of sources in the halo
region with the CCEDT task. These components were subtracted from the 
D array uv-data, and we used resulting data to produce a low resolution image
of the diffuse component (see Fig. \ref{fig:2}). 
We used this image to measure the halo flux density and size.
The main problems in this procedure are the following:
1) The presence of extended emission from a head-tail source located near 
the center of the radio halo. We compared the flux density of this source
with the clean components used to
subtract it from D-array UV data. We are confident to have subtracted
more than 95$\%$ of the flux density of this extended tail source.
2) The elongated halo extension in the extreme east region. In this position, 
two discrete sources are present, one coincident with a X-ray source. 
However, the residual flux density in the low 
resolution image is too high to be interpreted as the residual of the 
subtracted sources. In any case, including this region or not
will change the halo flux density by only $\sim$ 2.7 mJy.

The estimated halo flux density is 59 $\pm$ 5 mJy, where the error includes
the uncertainties in the source subtraction), corresponding to a total
radio power = 1.47 $\times$ 10$^{24}$ W/Hz (Log P = 24.17). The morphology 
is irregular with a largest 
angular size of $\sim$ 12 arcmin 
in the EW direction, corresponding to $\sim$ 1.3 Mpc.

The radio emission detected at high resolution and shown in Fig. 1 
permeates both merging clumps, as found e.g. in the Bullet cluster 
(\cite{lia00}). This is also similar to the case e.g. of A2255 (\cite{gov05})
or A665 (\cite{gio00}) where the radio halos are elongated in the direction 
of the cluster merger. The radio structure in A523 is therefore typical of 
radio halos. When degraded to lower resolution, owing to the improved
signal-to-noise ratio, the low brightness radio emission is more extended in 
particular
in the E-W direction. The elongated structure does not show any morphological 
feature typical of
radio relics such as high brightness filamentary structures 
or a transverse flux
asymmetry (see e.g.\cite{vwe11}). We conclude that its classification 
as a radio 
relic is very unlikely. We also note that a fainter X-ray emission, not 
detected by ROSAT, could be present.
We therefore classify the diffuse radio emission in A523 as a giant radio halo.
The size and radio power of the halo are in good agreement with the 
correlation presented for radio halos
by \cite{gio09}.

\subsection{Optical and X-ray data}

\begin{figure}
\centering
\includegraphics[width=6.5cm]{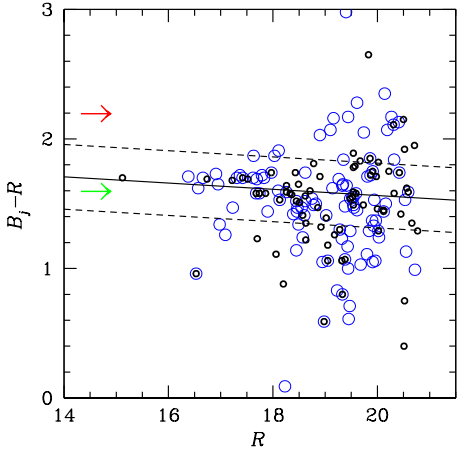}
\caption
{$B_j$--$R$ vs. $R$ diagram for galaxies extracted from the SSS. Small
  circles indicate galaxies within a radius of 4\arcm from the BCG and
  used to fit the CMR (solid line).  The dashed lines are drawn at
  $\pm$0.25 mag from the CMR to delimitate the region of ``likely
  cluster members''. Large blue circles indicate galaxies within a
  radius of 4\arcm from the SSW peak. The green and red arrows roughly
  indicate predicted colors for early-type galaxies at $z=0$ and
  $z=0.2$, respectively.}
\label{fig:3}
\end{figure}

\begin{figure}
\includegraphics[width=6.5cm]{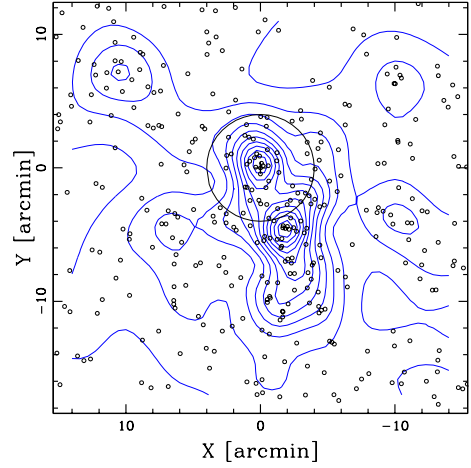}
\caption
{Spatial distribution on the sky and relative isodensity contour map
  of likely cluster members with $r^{\prime}\le 20$, obtained with the
  2D DEDICA method. The large circle indicates the region of
  4\arcm-radius around the BCG used to fix the CMR.}
\label{fig:4}
\end{figure}

Optically, A523 is a very rich cluster of Abell richness class $=2$
(Abell et al. \cite{abe89}). It was classified to be of the Rood-Sastry
morphological type ``L'', i.e. to have a linear configuration of
galaxies (Struble \& Rood \cite{str87}). To date, only the redshift
for the brightest cluster galaxy [BCG at
R.A.=$04^{\mathrm{h}}59^{\mathrm{m}}12\dotsec97$, Dec.=$+08\degree
49\arcmm 41.3\arcs$ (J2000.0)] has been acquired ($z=0.1036$; Crawford et
al. \cite{cra99}). No radio emission is clearly associated with this 
galaxy.
The visual inspection of DSS2-red image shows the presence of two
clumps. To quantify this impression, we compiled a photometric catalog
in a large region around the BCG ($\sim 30\arcmm$) from the SuperCOSMOS
Sky Surveys (SSS\footnote{http://www-wfau.roe.ac.uk/sss/index.html})
extracting the objects classified as ``galaxies'' that have both $B_j$
and $R$ magnitudes available.  In this
photometric catalog, we selected likely cluster members on the basis of
the color--magnitude relation (hereafter CMR), which indicates the
early-type galaxy locus. To determine the CMR, we considered the 75
galaxies within a radius of 4\arcm from the BCG and applied a
two-sigma-clipping fitting procedure, obtaining $B_j$--$R$=
2.044-0.024$\times R$ for a subsample of 54 galaxies. The color-magnitude
diagram is shown in Fig.~\ref{fig:3}.
From the photometric catalog, we considered as ``likely cluster
members'' those galaxies lying within 0.25 mag of the
CMR. Fig.~\ref{fig:4} shows a zoomed region of the contour map for
the likely members (734 galaxies within the whole $\sim$ 30\arcm
region).  The galaxy distribution reveals two significant peaks $\sim
5\arcm$ ($\sim 0.5$ Mpc) far: the NNE peak, coincident with the BCG,
and the SSW peak, with no dominant galaxy, at
R.A.=$04^{\mathrm{h}}59^{\mathrm{m}}05\dotsec4$, Dec.=$+08\degree
45\arcmm 06\arcs$ (J2000.0). 
The SSW peak has the higher density and a
larger population -- by factors of 1.2 and 2.7 according to the 2D
adaptive--kernel method (2D
DEDICA, Pisani \cite{pis96}).
Figure~\ref{fig:3} also shows the 105 galaxies of the photometric
sample  within a radius of 4\arcm from the SSW peak, whose properties can be
accurately described by the above CMR, too, suggesting that both clumps
are at a similar redshift.

 \begin{figure}
 \centering
 \includegraphics[width=7.0cm]{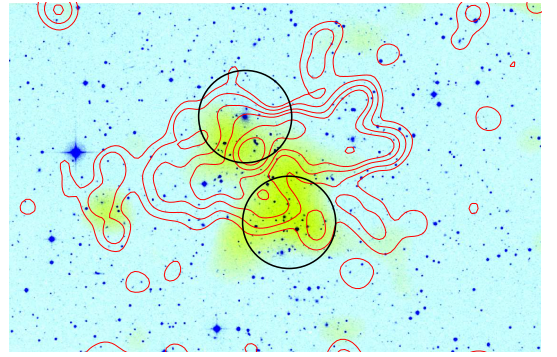}
    \caption{Multiwavelength picture of A523. A smoothed Rosat image 
(green color) is superimposed on the
DSS2-red image. The black circles indicate the regions of
2\arcm-radius around the NNE and SSW galaxy groups. The BCG is clearly
visible at the center of the NNE circle.  Contour levels of
the radio image of the extended halo after the subtraction of discrete
sources are shown in red.}
       \label{fig:5}
 \end{figure}

\cite{ebe98} report a cluster X-ray luminosity in the
0.1-2.4 keV band of 1.07 $\times$  10$^{44}$ erg/sec (after cosmological 
corrections). 
\cite{boe00} report a slightly lower value of 0.9 $\times$ 10$^{44}$
erg/sec.
The hot ICM (see Fig. \ref{fig:1}) is clearly bimodal in agreement 
with the optical galaxy distribution. However, a clear shift between the
galaxies and gas distribution is present (see Fig. \ref{fig:5}) and the
BCG is not coincident with the X-ray peak. 
The SSW structure is more extended
and shows an irregular shape. The NNE clump is more compact and regular.
The bright source, visible in the X-ray map at 
R.A.=$04^{\mathrm{h}}59^{\mathrm{m}}37\dotsec7$, 
Dec.=$+08\degree
45\arcmm 50\arcs$ (J2000.0) is  identified  with a discrete
unrelated radio source.

We conclude that optical and X-ray data that we have presented indicate 
a merging cluster structure, where
the main cluster is identified with the SSW structure. This cluster as
shown by its irregular shape and lack of a dominant galaxy, was not in a
relaxed stage, and is now strongly interacting (major
merging) with a more compact cluster (the NNE structure) dominated by a bright
BCG. 

\section{Discussion}

\begin{figure}
 \centering
 \includegraphics[width=7.0cm]{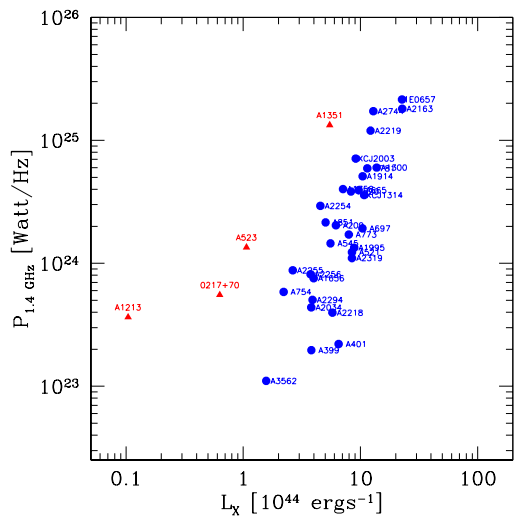}
    \caption{Cluster X-ray luminosity between 0.1 and 2.4 keV versus
the total halo radio power at 1.4 GHz. Red triangles are 
outlier clusters discussed in the text. Blue dots are 
classical clusters from \cite{gio09}.}
       \label{fig:6}
 \end{figure}

We present in Fig. \ref{fig:6} the plot of  
the total radio power and X-ray
luminosity for radio halos as shown by \cite{gio09}, where we have included
0217+70 and A523. 
The dots refer to classical powerful radio halos in
X-ray luminous clusters, which have been found to show a correlation between
the radio power and the X-ray luminosity. The red triangles are 
outliers, i.e. they refer to the few known radio halos that are 
overluminous in radio with
respect to the empirical radio - X-ray correlation. 

We focus here on the objects that are overluminous in radio with
respect to the empirical radio - X-ray correlation. We note
that there might be some uncertainties in the computation of their
parameters.  In particular, A1213 displays a peculiar diffuse emission, and its
size and radio power is lower than that of classical giant radio
halos.  Even if \cite{gio09} showed that small-size radio
halos have the same properties as giant radio halos, data of higher statistical
quality are necessary to confirm this result.  In addition, 
the cluster 0217+70,
as discussed by \cite{bro11}, is on the galactic plane therefore
the X-ray luminosity could be affected by an unusually high absorption.
On the other hand, the uncertainties in the determination of the radio
power and the X-ray luminosity in the cluster A1351 (\cite{gio09}) 
are smaller than the deviations from
the correlation (however see \cite{gia09}).

In the case of A523, we are aware that the redshift is 
not well known, since only the redshift of the BCG is available. However,
to be consistent with the correlation between the radio power and the 
X-ray luminosity, the
clusters should be at a redshift of $\sim$ 0.2. 
Fig.~\ref{fig:3}
shows the predicted values of the $B_j$--$R$=1.4 and 2.0  for the
predicted early-type galaxies at $z=0.0$ and $z=0.2$, respectively
(see Fig.~1 of Brown et al. \cite{bro00} and refs. therein) after a 0.2 mag 
decorrection for the Galactic absorption. Although these predictions
cannot be used to provide a redshift estimate, they are useful
to show to the reader the difference expected in the color--magnitude
relation for a $z\sim0.2$ cluster, thus strongly favouring the location of
A523 at much lower distance.
Moreover, at z = 0.2, the halo size would be $\sim$ 2.4
Mpc, the largest radio halo known to date (see \cite{gio09}).

The radio halo in A523 therefore represents a robust case of a
giant radio halo clearly identified with an X-ray underluminous cluster. 
As discussed in Sect. 2, the X-ray and optical 
distribution is similar (bimodal)
but not in perfect agreement. The thermal gas distribution is shifted
with respect to the galaxy distribution; this is more evident for the
smaller and more compact NNE clump, where the position of the BCG
is not at the center of the gas distribution.
A higher quality X-ray image is necessary to investigate 
whether the peculiarity of A523 
could be related to the possibility that the compact 
NNE group just crossed the SSW cluster as in the case of the Bullet cluster
(1E0657-56, \cite{mar02}), 
and this strong and recent interaction affected in a different way 
the gas and galaxy distribution.

We conclude that there is firm observational evidence that giant radio
halos can be associated with low luminosity X-ray clusters. These 
halos are overluminous in radio by at least an order of
magnitude with respect to that expected from the extrapolation of the
observed radio power - X-ray luminosity.

The radio - X-ray correlation is consistent with reacceleration
models, which have been extensively discussed in the literature
(Cassano 2010 and refs therein) and supported by the observations
(e.g. \cite{gio09}). In the framework of these models, the radio
emitting particles in cluster radio halos gain energy from
merger-driven turbulence in the ICM.  Turbulence
development on timescales of the order of 1 Gyr is necessary to
reaccelerate electrons to the energies needed to emit the observed
synchrotron radiation at GHz frequencies, thus giant radio halos are
expected to be strictly connected to massive systems that have
undergone strong merging processes.
In the framework of turbulent reacceleration models, Cassano 2010 suggests 
that very steep spectrum radio halos should exist that are detectable 
only at low frequencies, and be less luminous than predicted 
by the radio - X-ray correlation. The radio halo in A523, and a
few similar objects, are instead more luminous in radio than predicted by the
radio - X-ray correlation.
These powerful radio halos, associated with clusters of low X-ray 
luminosity, do not appear to be described well by current models, hence
are likely to represent a new class of objects, raising new questions
about the origin of radio halos.  They could be either young halos 
or clusters at a special
time of the merger event, when
particle acceleration processes have a higher efficiency 
(see \cite{bru11}). Another possibility is that the
X-ray luminosity might not be in these cases a good indicator of the previous
cluster merging activity. 

We note that the non-thermal diffuse radio emission in A523 
does not follow the optical and the X-ray emission, unlike most 
other radio halos.
The brighter halo region visible at high 
resolution (Fig. 1) is in good agreement with the X-ray bimodal 
distribution, 
but the more extended low brightness halo emission, which is most clearly 
imaged at low resolution, is elongated in the E-W direction
i.e. in the direction perpendicular to the merger. This morphology could 
be related to peculiar cluster conditions that give rise to these overluminous
radio halos and needs to be studied in better detail with more sensitive
X-ray and radio data and radio spectral information.

\begin{acknowledgements}
The National Radio Astronomy
Observatory is operated by Associated Universities, Inc., under cooperative
agreement with the National Science Foundation.
\end{acknowledgements}

\end{document}